# Features of spherical torus p-$^{11}$B burning plasmas

**Y.-K. M. Peng, A. Ishida, T. Sun, W. Liu, H. Huang, Y. Shi, B. Liu, D. Guo, Z. Li, D. Luo, X. Xiao, G. Zhao, and M. Liu**

ENN Science and Technology Development Co Ltd, Langfang, China

Email: pengyuankai@enn.cn; pengykm2@gmail.com

---

## Abstract

A spherical torus (ST) p-$^{11}$B plasma model satisfying a multi-component magnetofluid force balance is developed, which includes small fractions of suprathermal ions with characteristic temperatures around 0.5 MeV and suprathermal electrons at energies in the MeV range. Alongside the primary thermal plasma with ion temperatures exceeding 100 keV and densities above $10^{20}$ m$^{-3}$, these components raise fusion reaction rates by exploiting the p-$^{11}$B fusion cross section, which exhibits a local peak near 160-165 keV and a broader, higher peak near 650-675 keV [1,2]. Suprathermal ions and strong toroidal rotation driven by neutral beam injection (NBI) have been observed in devices such as START, MAST, NSTX, Globus-M2, and ST40 [3-8], achieving high energy confinement. Central-solenoid-free plasma initiation, ramp-up, and sustainment were designed and tested on EXL-50 [9] and replicated on EXL-50U with partial central induction [60,61,10], demonstrating that relativistic electrons enable efficient drive current, consistent with the multi-magnetofluid equilibrium model [11]. Motivated by ENN's aneutronic commercial fusion roadmap [12], this paper presents a rotating, thermally non-equilibrium ST p-$^{11}$B force-balance model with several distinctive properties: fluid components experience separate balance under centripetal, electrostatic and Lorentz forces with common electric and magnetic fields, producing large rotation-speed differences between thermal boron ions and suprathermal protons; a large outboard region with magnetic well and omnigeneity is formed, altering neoclassical transport and gradient-driven turbulences; suprathermal charged particles can extend beyond the last closed flux surface (LCFS) and be limited by plasma-facing components, affecting recycling and pedestal conditions; and superposition of plasma components modifies sources and sinks of free energy, motivating renewed evaluation of stability, turbulence, transport, heating, current drive, and flux diffusion. Challenges and opportunities for sustained burn are discussed for a compact ST p-$^{11}$B plasma with 1.4-m major radius, 13-MA current, and 3-T toroidal field.

---

## 1. Introduction

ENN has adopted a commercial development roadmap for aneutronic p–$^{11}$B fusion power with a goal of demonstrating fusion electricity production by 2035 [12]. The spherical torus (ST) has been selected as the primary magnetic-confinement configuration in this roadmap. The motivation is that ST plasmas can exhibit high beta, strong toroidal rotation, and favorable confinement properties at compact size, and may provide access to regimes where suprathermal components contribute materially to fusion performance [3–8].

### 1.1. Challenges facing scientific break-even of p–$^{11}$B plasmas

The potential advantages of aneutronic p–$^{11}$B fusion have attracted attention since early studies of advanced fuels [13]. McNally emphasized that, for thermonuclear p–$^{11}$B plasmas, bremsstrahlung radiation can exceed fusion power under broad conditions, imposing a severe power-balance constraint [13]. Rider identified fundamental limitations for plasma fusion systems that are not in thermodynamic equilibrium, concluding that maintaining non-Maxwellian electron distributions to offset electron–ion collisional coupling requires prohibitive input power in many settings [14].

Nevins reviewed confinement requirements for advanced fuels [15] and, with Swain, provided formulations of the thermonuclear p–$^{11}$B fusion rate coefficient [2]. These works clarified that approaches relying on a reduced electron temperature relative to the ion temperature and an increased effective charge can reduce bremsstrahlung relative to fusion power but remain constrained by collisional energy exchange and the need for sustainment power. Sikora and Weller [16] re-evaluated the $^{11}$B(p, α)αα reaction rates and argued for the possibility of substantially larger effective cross sections, which could bring bremsstrahlung losses closer to balance with fusion power under idealized assumptions for uniform, unmagnetized plasmas. Tentori and Belloni revisited the p–$^{11}$B cross section and reactivity and provided analytic approximations useful for plasma reactivity modelling [1].

### 1.2. Motivation for suprathermal ions

A long line of magnetic-confinement research has established the importance of suprathermal-plus-thermal ion components for enhancing reactivity in D–D and D–T plasmas. Jassby proposed beam-driven toroidal reactor concepts using a suprathermal beam in a thermal plasma target [17,18], and such ideas were part of the physics basis for high-power tokamak experiments, including TFTR [19]. In the context of advanced fuels, Rostoker, Binderbauer, and Monkhorst proposed suprathermal colliding p–$^{11}$B beams as a path to avoid heating the entire plasma to the several-hundred-keV range [20]. However, sustaining suprathermal populations can require large input power, preventing attainment of $Q_{sci} > 1$ in colliding-beam scenarios [21].

Putvinski et al. [22] re-assessed p–$^{11}$B fusion reactivity and evaluated the potential of elastic nuclear collisions between fusion alphas and thermal protons to enhance reactivity through avalanche-like multiplication, indicating a possible improvement at the ~10% level under idealized conditions. Complementary work raised the prospect of avalanche proton–boron fusion based on elastic nuclear collisions [23] and prompted critical discussion of its magnitude [24]. Belloni further analyzed multiplication processes in high-density H–$^{11}$B fusion fuel [25], and Tentori and Belloni clarified how chain-reaction-type amplification involving fusion alphas and thermal protons may operate, linking such mechanisms to beam–plasma target concepts and to laser-driven p–$^{11}$B schemes [1,26]. Progress in laser-driven p–$^{11}$B fusion tests has been reviewed by Vovkivsky and Chirkov, including reports of very large fusion-alpha yields and discussion of reduced stopping and increased effective interaction depth [27]. Analogous ideas in magnetic confinement include the use of RF wave interactions to modify suprathermal-ion distributions, as discussed for JET DTE2 [28], and alpha-channeling concepts to improve ion-heating efficiency [29,30].

### 1.3. Suprathermal components in ST p–$^{11}$B plasmas

The p–$^{11}$B fusion cross section has been studied since early measurements by Oliphant and Rutherford [31] and by Dee and Gilbert [32]. A more complete theoretical model of reaction channels was developed by Quebert and Marquez [33]. These nuclear-physics results motivated proposals for spin-polarized fuels; Kulsrud et al. discussed the physics of spin-polarized plasmas [34], and the effects of polarization on the p–$^{11}$B cross section were quantified by Dmitriev [35] and by Ahmed and Weller [36]. Tokamak-based experimental research has been described to test polarization survival and related issues [37]. Spin polarization in p–$^{11}$B plasmas in the ST context remains open for initial theoretical exploration.

ST experiments have demonstrated regimes with suprathermal ions and strong toroidal rotation driven by NBI. Such behavior has been reported for START and related work [3,4], NSTX high-beta rotating plasmas [5], MAST confinement and transport studies [6], Globus-M2 high-field performance with large increases in triple product [7], and ST40 high-field compact operation with central ion temperatures reaching $10^8$K [8]. These results support the premise that compact, strongly rotating ST plasmas can access the high-performance regimes required for advanced fusion fuel such as p-$^{11}$B.

In addition, central-solenoid-free plasma initiation, current ramp-up, and sustainment via ECRH were developed and tested on EXL-50 [9] and extended to EXL-50U, where 1 MA-class discharges in hydrogen–boron plasmas have been reported [10]. These scenarios rely on a percent-level population of suprathermal electrons with relativistic energies in the 0.2 – 1.0 MeV range. The high-current-drive effectiveness of such relativistic electrons is consistent with equilibrium reconstructions based on multi-component magnetofluid force balance [11].

### 1.4. Strategy and scope of this paper

The p – $^{11}$B cross section exhibits a local peak near 160 – 165 keV and a broader peak near 650 – 675 keV [1,2]. The main plasma therefore comprises thermal protons and $^{11}$B ions above 100 keV at densities above $10^{20}$ m$^{-3}$, with minority suprathermal ions near 0.5 MeV. Rostoker et al. proposed suprathermal reactants with an average energy separation of ~580 keV [20]. A multi-MeV suprathermal electron component is included to represent multi-harmonic ECRH consistent with the EXL-50/EXL-50U data [9,10].

The requirements for sustained advanced-fuel burn can be organized by increasing physical timescales spanning equilibrium force balance, orbit containment of suprathermal charged particles, macroscopic stability, microturbulence and transport, current drive and magnetic flux diffusion, plasma – wall interaction, and material erosion. The present paper focuses on the first two—equilibrium and orbit physics, including the effects of large suprathermal proton orbits on the fusion reaction rate—because these set the operating conditions for subsequent stability and transport studies.

Section 2 reviews the axisymmetric multi-magnetofluid equilibrium model [11], contrasting it with the MHD equilibrium reconstruction [38] and with operational-limit considerations [39]. Section 3 extends the model to ST p – $^{11}$B plasmas approaching burn and provides a computational approach that yields numerical solutions for macroscopic quantities with 1% numerical accuracy. Section 4 shows that an orbit model is required to compute suprathermal proton fusion reaction rates with thermal $^{11}$B ions, because energetic-ion drift orbits can deviate substantially from local-flux-surface assumptions; related finite-orbit-width effects have been included previously for energetic beam ions [40], and orbit concepts have been used to interpret current profiles driven by cyclotron-heated suprathermal electrons in EXL-50 [41]. Section 5 discusses confinement and stability implications in contrast with ITER-like thermonuclear plasmas [42,43] and standard neoclassical transport theory [44], motivating validation in the EXL-50U and upcoming EHL-2 ST experiments [45,46].

---

## 2. Multi-magnetofluid equilibrium database and application to ST p-boron plasmas

### 2.1. Motivation and early applications to ST plasmas

Rotation and suprathermal populations can modify equilibrium force balance in ways not captured by single-fluid ideal-MHD reconstructions. A multi-component approach has therefore been used to study the role of toroidal rotation and energetic particles in low-aspect-ratio equilibria, including ST plasmas sustained by RF and ECRH.

The development of the EXL-50 p‑boron experiment required an axisymmetric equilibrium description that includes multiple fluids, and particularly a relativistic suprathermal electron component. A four-fluid axisymmetric plasma equilibrium model including relativistic electrons was developed, together with a computational method enabling reconstruction of available experimental data and prediction of macroscopic profiles consistent with multi-magnetofluid force balance [11]. A particle-orbit model for cyclotron-driven energetic electrons was introduced to visualize the non-local nature of current formation in such plasmas [41]. The orbit picture indicates that the driven-electron-current density can peak near the outboard LCFS even when the suprathermal electron pressure is centrally peaked, as the energetic electron orbits can extend beyond the LCFS. Since suprathermal protons at energies of a few hundred keV to MeV can also have orbit widths that are not small compared with equilibrium gradient scale lengths, a non-local approach is needed when computing the p‑$^{11}$B reaction rate contribution from suprathermal ions (see Section 4).

### 2.2. EXL-50U plasma equilibrium under ECRH alone (reconstruction with relativistic electrons)

We summarize an extension of the multi-magnetofluid equilibrium model to reconstruct recent EXL-50U discharges achieved using ECRH alone, in the presence of a central solenoid (CS) coil [10]. The reconstructed discharge has plasma current $I_p \approx 0.8$MA (shot #10237 at $t \approx 0.45$s). The equilibrium is computed by imposing force balance for four magnetofluids: thermal protons, boron ions (natural boron), thermal electrons, and a relativistic suprathermal electron component [11]. Using this framework, all available experimental constraints are reproduced within a few percent. Direct profile measurements for relativistic electrons are not yet available; therefore, their inferred profiles are model outputs constrained to be consistent with macroscopic equilibrium force balance and with the plasma measured parameters that serve as the reconstruction targets.

A central inference from the reconstruction is that approximately 90% or more of the total plasma current is carried by the relativistic suprathermal electrons. This partition is not a direct measurement, but model results consistent with the available equilibrium constraints; it is also compatible with the non-local orbit picture in which energetic-electron trajectories can extend beyond the LCFS [11,41]. For this case, the computed LCFS is separated by about 7 cm from the outboard limiting plasma-facing material, and the toroidal current associated with the relativistic electrons extends beyond the LCFS both radially and vertically.

This separation between LCFS and limiter, together with non-local electron orbits, is a defining ingredient for interpreting the inferred current distribution and for the expectation that edge interaction and recycling may differ from those in the conventional tokamaks.

Figure 1 summarizes the reconstructed equilibrium structure: (a) poloidal flux contours, and (b) the toroidal current density profile $J$. The red curve denotes the LCFS; the black curve marks the boundary of the relativistic-electron population; and the blue curve indicates the limiting wall.

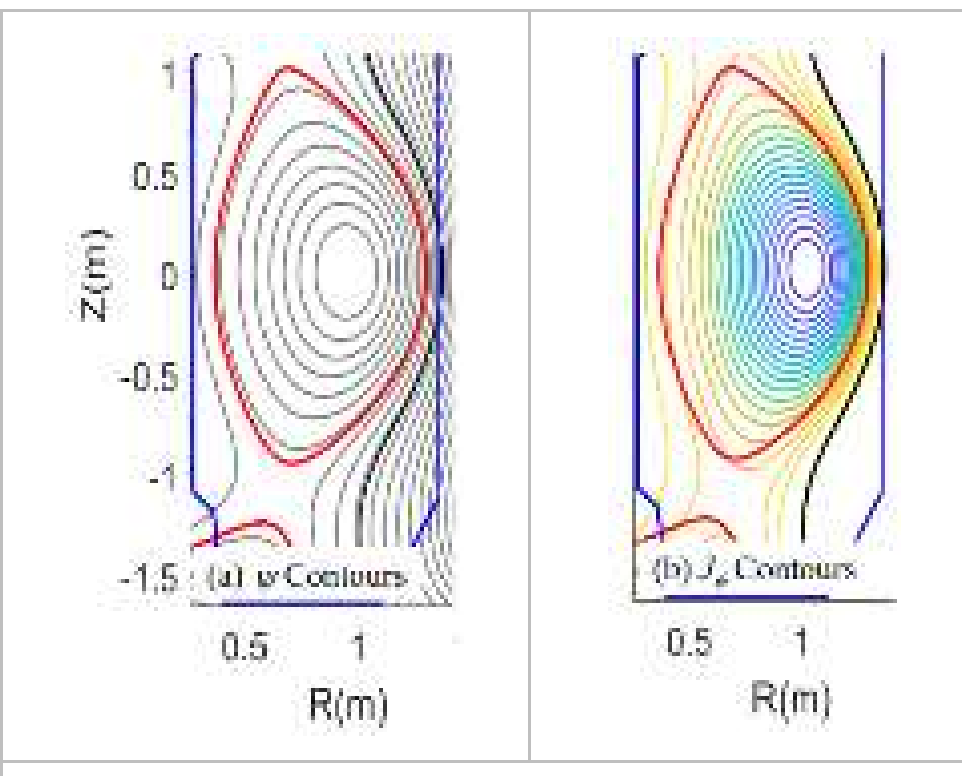

**Fig. 1.** (a) Contours of poloidal flux, and (b) toroidal current density profile $J_\phi$. The red line indicates the LCFS, the black line shows the boundary for the relativistic electrons, and the blue line marks the limiting plasma-facing material.

Figure 2 presents reconstructed mid-plane profiles of (a) toroidal current densities, (b) densities, (c) toroidal rotation velocities, and (d) temperatures for boron, proton, thermal electrons, and relativistic suprathermal electrons (subscripts b, p, el, and eh, respectively). For the relativistic-electron component, direct profile measurements are not yet available; the profiles shown should therefore be interpreted as equilibrium-consistent model outputs conditioned on the reconstruction targets and assumptions [11]. In ECRH-only operations without direct ion heating, relativistic electrons dominate several key macroscopic quantities. This behavior is expected to change as stronger ion heating is added to EXL-50U and the upcoming EHL-2 experiment, which is currently under preparation [45,46].

## 3. Extension to p - $^{11}$B burning plasmas

### 3.1. Formulation for thermal plus suprathermal p - $^{11}$B plasma

We extend the multi-magnetofluid equilibrium framework of Ishida et al. [11] to projected ST p - $^{11}$B plasmas as they approach a sustained burn. The component set used includes thermal protons (pl), suprathermal protons (ph), thermal $^{11}$B ions (bl), suprathermal $^{11}$B ions (bh), thermal electrons (el), and suprathermal electrons (eh). Since the suprathermal electron fluid can be relativistic, relativistic continuity and momentum equations are used for that component.

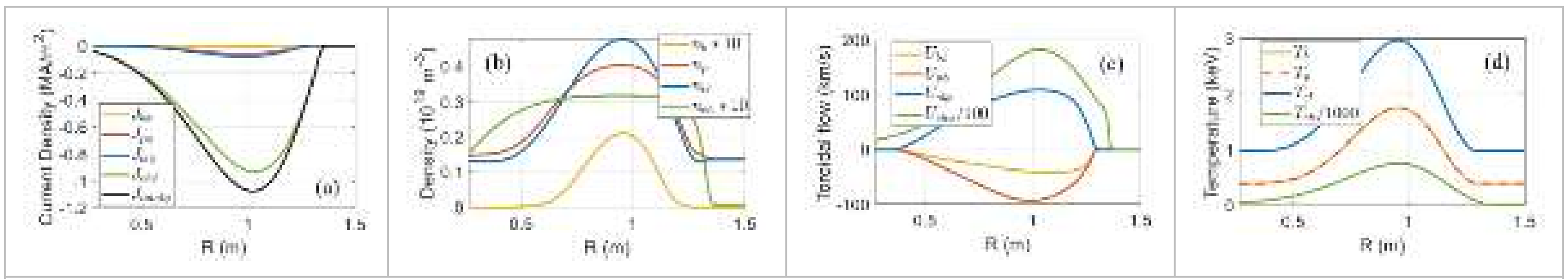

**Fig. 2.** Mid-plane profiles of (a) toroidal current densities, (b) densities, (c) toroidal rotation velocities, and (d) temperatures of boron, proton, thermal electrons, and relativistic suprathermal electrons identified by subscripts: b, p, el, and eh, respectively.

For each component fluid:

$$\frac{\partial(\gamma n)}{\partial t} + \nabla \cdot (\gamma n \mathbf{u}) = 0, \qquad (1)$$

$$\frac{\partial}{\partial t}\left(mn\gamma^2 g_{ep}(T^*)\,\mathbf{u}\right) + \frac{\partial}{\partial x_j}\left(mn\gamma^2 g_{ep}(T^*)\,\mathbf{u}u_j\right) + \nabla p = q\gamma n\mathbf{E} + q\gamma n\,\mathbf{u}\times\mathbf{B}. \qquad (2)$$

Here $p = nT$, $\gamma = (1 - u^2/c^2)^{-1/2}$, and $g_{ep}(T^*) = (\epsilon + p)/(mnc^2)$, with $T^* = T/(mc^2)$. The equilibrium force-balance equations are:

$$\nabla \cdot (\gamma n \mathbf{u}) = 0, \qquad (3)$$

$$m\gamma\mathbf{u}\cdot\nabla\left(\gamma\mathbf{u}\,g_{ep}\right) + \nabla T + T\nabla \ln\, n + q\gamma\nabla V_E = q\gamma\mathbf{u}\times\mathbf{\Omega}, \qquad (4a)$$

$$\mathbf{\Omega} \equiv \mathbf{B} + \nabla\times\left(q^{-1}m\gamma g_{ep}\mathbf{u}\right), \qquad (4b)$$

with $V_E$ being the electrostatic potential.

In cylindrical coordinates $(R, \phi, Z)$, we define:

$$\mathbf{B} = \nabla\psi(R,Z)\times\nabla\phi + RB_\phi\nabla\phi, \qquad (5a)$$

$$\mathbf{\Omega} = \nabla Y(R,Z)\times\nabla\phi + R\Omega_\phi\nabla\phi, \qquad (5b)$$

$$\gamma n\mathbf{u} = \nabla\Phi(R,Z)\times\nabla\phi + R\gamma n u_\phi\nabla\phi, \qquad (5c)$$

where $Y = \psi + (m/q)\gamma g_{ep} R u_\phi$ and $\Omega_\phi = B_\phi - R\nabla\cdot\left((mg_{ep})/(qnR^2)\,\nabla\Phi\right)$.

Axisymmetry requires $\Phi = \Phi(Y)$. Adopting $\mathbf{\Omega}\cdot\nabla T = 0$ gives $T = T(Y)$ and hence $g_{ep} = g_{ep}(T(Y)/mc^2)$. Taking the component of equation (4a) along $\mathbf{\Omega}$ leads to $F = F(Y)$, with

$$F \equiv \frac{1}{2}m(\gamma u)^2 g_{ep} + T(1+\ln\, n) + qV_E. \qquad (6)$$

Thus, each component fluid is specified by three profile functions $F(Y)$, $T(Y)$ and $\Phi(Y)$, with $Y = RP_\phi/q$. The toroidal flow can be written as

$$\gamma u_\phi = \frac{R}{q}\left(\frac{dF}{dY} - \frac{dT}{dY}\ln\, n + \frac{1}{2}m(\gamma u)^2\frac{dg_{ep}}{dT}\frac{dT}{dY}\right) + \frac{1}{n}\frac{d\Phi}{dY}\Omega_\phi. \qquad (7)$$

Ampère’s law becomes

$$R\frac{\partial}{\partial R}\left(\frac{1}{R}\frac{\partial\psi}{\partial R}\right) + \frac{\partial^2\psi}{\partial Z^2} = -\mu_0 R\sum_\alpha q_\alpha\, n_\alpha\gamma_\alpha u_{\alpha\phi}, \qquad (8a)$$

$$RB_\phi = \sum_\alpha \mu_0\, q_\alpha\Phi_\alpha. \qquad (8b)$$

Here, $\alpha \in \{pl, ph, bl, bh, el, eh\}$. In the regimes considered, the poloidal mass flow is much smaller than the toroidal mass flow, so $\Omega_\phi \approx B_\phi$, and the remaining term is treated as a correction. The numerical solution method follows that of [11], using a three-grid scheme and SOR on a $100\times100\,(Z\times R)$ mesh. A $100\times200$ mesh case confirms the mesh-size convergence at about the 1% level. The electrostatic potential is substituted using equation (6), and four coupled density-ratio equations are solved iteratively to obtain a self-consistent $V_E$ [11].

### 3.2. Numerical iterative convergence logic

Dimensionless variables are formed using a reference length $L_{ref}$, a reference current $I_{ref}$, and a reference density $n_{ref}$. Derived reference quantities include

$$B_{ref} \equiv \frac{\mu_0 I_{ref}}{L_{ref}},\ u_{ref} \equiv \frac{B_{ref}}{\sqrt{\mu_0 m_p n_{ref}}},\ T_{ref} \equiv m_p u_{ref}^2,$$

and

$$\varepsilon \equiv \frac{c}{L_{ref}\sqrt{e^2 n_{ref}/(\varepsilon_0 m_p)}}.$$

A dimensionless set of profile functions in hyperbolic-tangent form provides flexibility while yielding convergent solutions with experimentally plausible profiles. Boundary conditions assume $V_E = 0$ on the computational outboard boundary, where $\Phi_\alpha = 0$ is adopted. A model flux function is used:

$$\psi_{model} = \psi_{pfcoil} + \psi_{plasma-model},$$

consistent with the EXL-50 and EXL-50U choices of poloidal flux boundary values [11].

An iterative convergence logic is adopted:

1. Choose input coefficients $C_{T(\alpha)0}$, $C_{F(\alpha)0}$, $C_{T(\alpha)1}$, $C_{F(\alpha)1}$ for each component fluid.
2. Set $Y_\alpha = \psi_{model}$ and $j_t = j_{t-model}$, or alternatively update using $Y_\alpha = \psi_{update}$ and $j_t = j_{t-update}$.
3. Solve Ampère's law to obtain $\psi$ with the specified boundary value; update $Y_\alpha$ and $u_{\alpha\phi}$.
4. Solve the four density-ratio equations iteratively.
5. Update $n_\alpha$.
6. If the convergence of Ampère's law is insufficient, return to step 2, and if the convergence of Ampère's Law is sufficient, verify that $V_E$ is computed accurately.
7. Check whether $(\psi - \psi_{plasma})$ is consistent with $\psi_{pfcoil}$ within the required tolerance; if not, return to step 2.
8. Stop when the above conditions are met; otherwise, return to step 1, and update the $\psi$ boundary value using the current solution.

### 3.3. Characteristics of high-$\boldsymbol{\beta}$ ST burning p-$^{11}$B plasmas

We compute example ST p-$^{11}$B equilibria approaching burn. A representative parameter set (units: m, MA, T, $10^{20}$ $\mathrm{m}^{-3}$, keV) is:

- $R = 1.43$, $a = 0.86$
- $I_p = 12.6$, $B_T = 3.38$
- $\beta_T = 0.25$, $\beta_p = 1.09$, $\beta_N = 5.9$
- $\kappa = 2.16$, $\delta = 0.27$
- $q_0 = 1.35$, $q_{LCFS} = 5.33$, $l_i = 0.28$

Central densities are:

- $n_{e0} = 1.04$, $n_{p0} = 0.43$, $n_{B0} = 0.12$
- $n_{ph0} = 0.0046$, $n_{Bh0} = 0.0014$, $n_{eh0} = 0.011$

Central temperatures are:

- $T_{p0} = T_{B0} = 132$, $T_{e0} = 66.7$
- $T_{ph0} = T_{Bh0} = 892$, $T_{eh0} = 2374$

Current partition is:

- $I_{thermal} = 9.89$, $I_{suprathermal} = 2.70$

This plasma composition is designed to take advantage of the sharp fusion cross-section spike near 160 - 165 keV and the broader, higher cross-section near 650 - 675 keV of the p - ¹¹B reaction [1,2]. The electron relativistic coefficient $g_{ep}$ in this case is 19.1 for suprathermal electrons and 1.35 for thermal electrons.

Figure 3 shows representative poloidal cross-section contours: (a) total toroidal current density, (b) total plasma pressure, (c) suprathermal-electron current density, (d) electrostatic potential, and (e) outboard magnetic-well and hill structure in $|B|$, indicating an axisymmetric omnigenous region consistent with early calculations of the spherical-torus plasma features [47], consistent with the broader omnigeneity concepts [48,49]. The computed suprathermal electron population extends beyond the LCFS and carries $I_{suprathermal}/I_p = 2.70/12.6 \approx 0.214$, i.e., about 21%, of the total plasma current, while the thermal current remains inside the LCFS. A positive electrostatic potential is obtained self-consistently; the peak value is 10.8 kV relative to the wall facing the plasma.

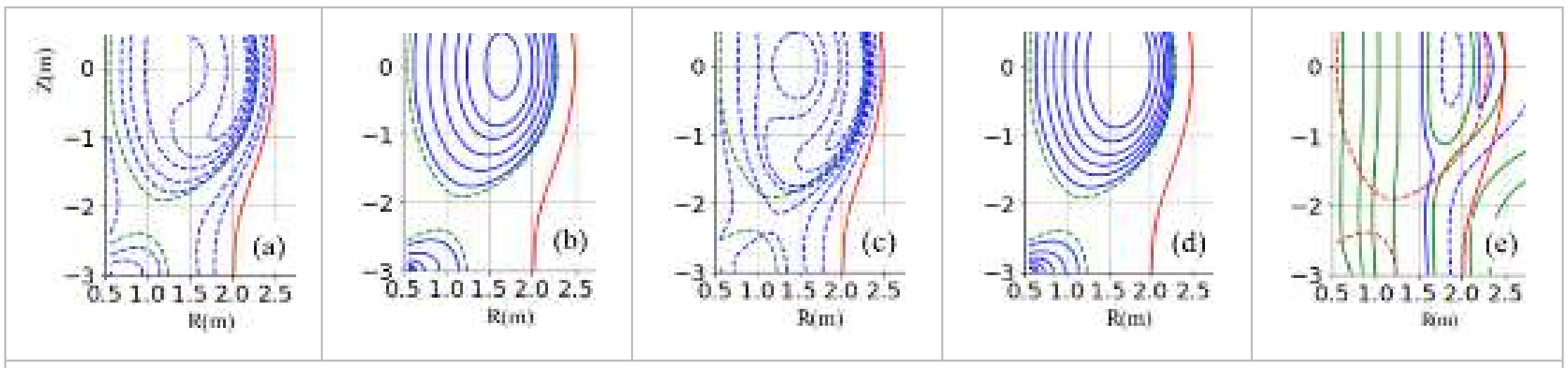

**Fig. 3.** (a) Toroidal current density (−0.1 to −4.0 MA m⁻²), (b) plasma pressure (0.1 to 2.6 MPa), (c) suprathermal electron current density (−0.01 to −0.5 MA m⁻²), (d) electrostatic potential (0.5 to 10.8 kV), and (e) outboard magnetic well (2.75 T) and hill (3.14 T) indicating an omnigenous region. The green or red dotted lines represent the LCFS; the red solid line represents the flux-surface boundary of the suprathermal components

The mid-plane rotation profiles are shown in Figure 4 for (a) protons, (b) boron, and (c) electrons, with thermal components in blue and suprathermal components in red, together with (d) the velocity difference between suprathermal protons and thermal boron ions over the poloidal cross section. Suprathermal protons rotate strongly near the LCFS, with a velocity reaching approximately 2700 km s $^{-1}$ relative to thermal boron ions. The velocity difference exceeds $2 \times 10^6$ m s$^{-1}$ over a substantial region, representing a significant fraction of the local suprathermal-proton thermal speed. Differential rotation between protons and borons has been proposed to increase the effective fusion reactivity under idealized conditions [50]. The outboard $|B|$ well suggests orbit squeezing that can reduce neoclassical ion transport, with improved scaling consistent with the magnetic-well transport modelling [51,52].

## 4. Orbit-model-based fusion reaction rates

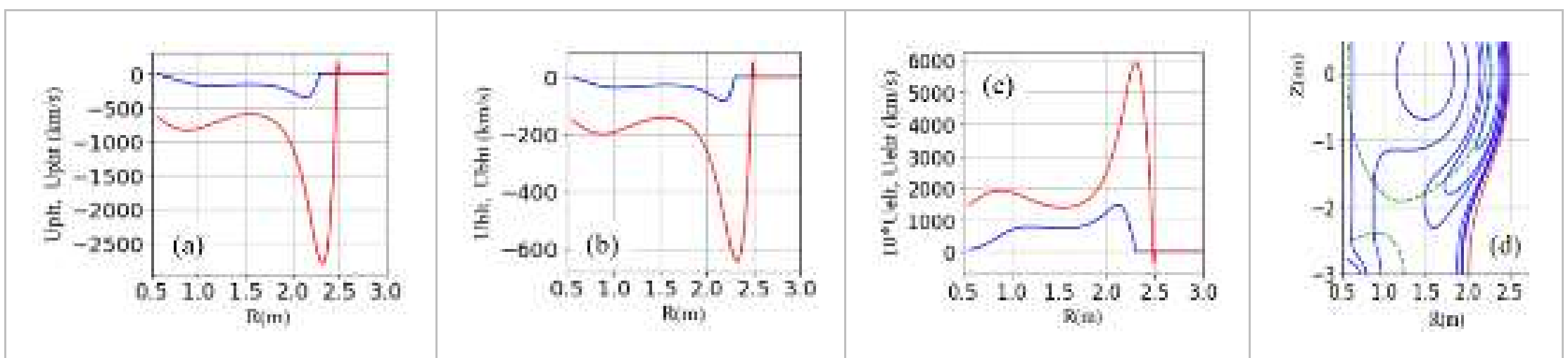

**Fig. 4.** (a) Proton, (b) boron, and (c) electron thermal (blue) and suprathermal (red) toroidal rotation velocities (km $s^{-1}$) on the mid-plane, and (d) the velocity difference between suprathermal proton and thermal boron (0.5 to 2.50 Mm $s^{-1}$) over the poloidal cross section. The green dotted line indicates the LCFS.

### 4.1. Zero-dimensional local model of fusion reaction rate and fusion power

Early experimental determinations of low-energy p‑$^{11}$B cross sections were developed from beam‑target measurements and remain a basis for evaluating local reactivity models [53]. The p‑$^{11}$B cross section has also been represented using theoretical modelling intended to capture reaction-channel behavior over relevant energies [54]. In magnetic confinement modelling, an idealization is to compute a local $\langle\sigma v\rangle$ from a velocity-space integral using only local ion parameters (densities, temperatures, and drifts). For the projected burning ST p‑$^{11}$B equilibrium of Section 3, we first apply such a local model using drifting Maxwellian distributions, following the drift bi-Maxwellian reactivity formulation [55].

A key question is whether the large rotational-velocity difference between suprathermal protons and slowly rotating thermal $^{11}$B ions (Figure 4) can raise the total fusion power density, as proposed in [50]. Figure 5 shows contours of p‑$^{11}$B fusion power density computed from the local-parameter model for (a) suprathermal protons reacting with thermal borons and (b) the remaining combinations of the proton boron components.

For the example equilibrium, the local-parameter model yields a volume-averaged fusion power density $P_{pB}$ of $0.0306\ \mathrm{MW\,m^{-3}}$. The suprathermal-proton/thermal-boron contribution is calculated to be 8.8% of the total, i.e., $0.0027\ \mathrm{MW\,m^{-3}}$. The corresponding total fusion power is $P_{pB,tot} = \int_{\mathcal{V}} p_{pB}(\mathbf{x})\, dV$, where $p_{pB}(\mathbf{x})$ is the local fusion power density based on the local $\langle\sigma v\rangle$ model and $\mathcal{V}$ denotes the plasma volume. These local-model results suggest that suprathermal protons contribute only a minor amount to fusion power. The remainder of this section explains why that conclusion can be incorrect for orbit-width-dominated suprathermal ions in a compact ST.

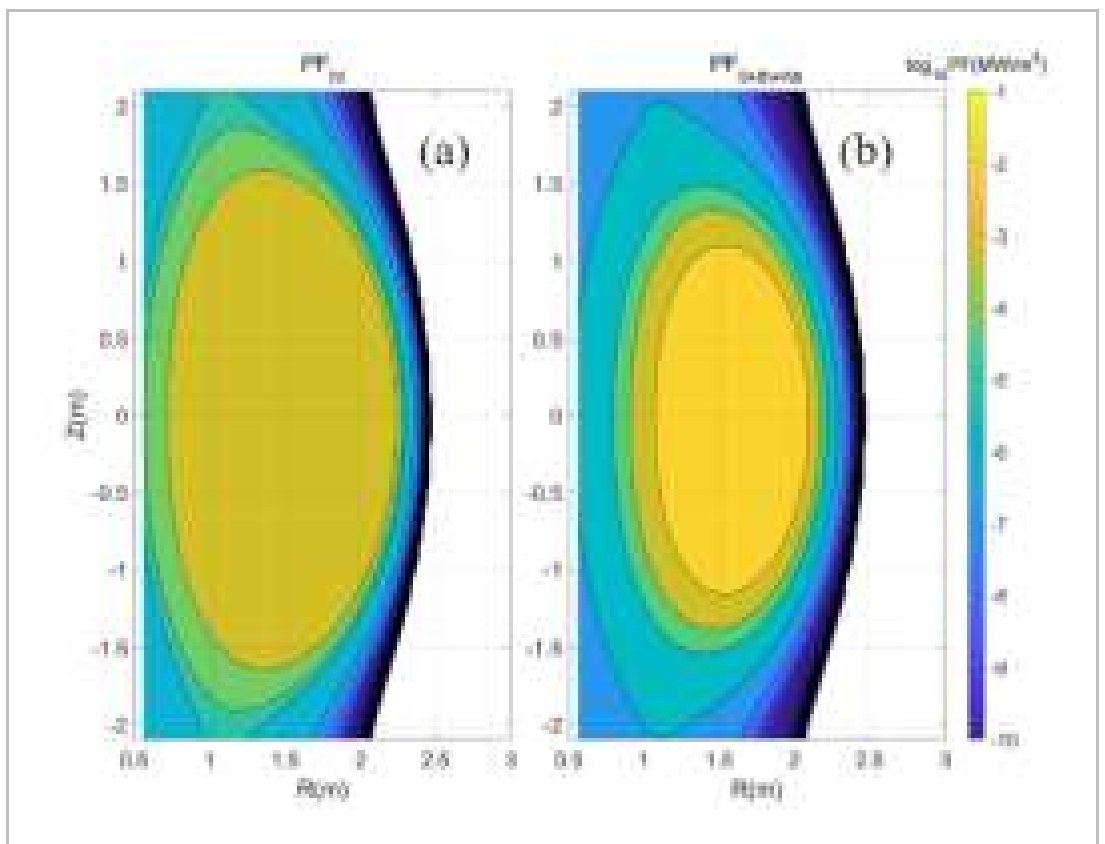

**Fig. 5.** Contours of $p-^{11}B$ fusion power density based on 0D local-parameter $\langle \sigma v \rangle$ integration in velocity space of (a) suprathermal protons with thermal borons, and (b) the remaining proton and boron component fluid combinations.

### 4.2. Non-local nature of suprathermal proton orbits

The main deficiency of the local-parameter approach is that it neglects non-local orbit physics of suprathermal protons in a low-aspect-ratio, high-beta, rotating equilibrium. Finite orbit widths of energetic ions have been incorporated previously in equilibrium models [40]. Orbit-based interpretations of current formation by electron cyclotron wave-driven suprathermal electrons in EXL-50 show strong non-local behavior relative to flux surfaces [41]. In the present case, suprathermal protons with MeV-level energies can traverse, in a single orbit, regions of substantially different thermal-boron densities, so the fusion probability cannot be calculated accurately from plasma parameters at a single spatial point.

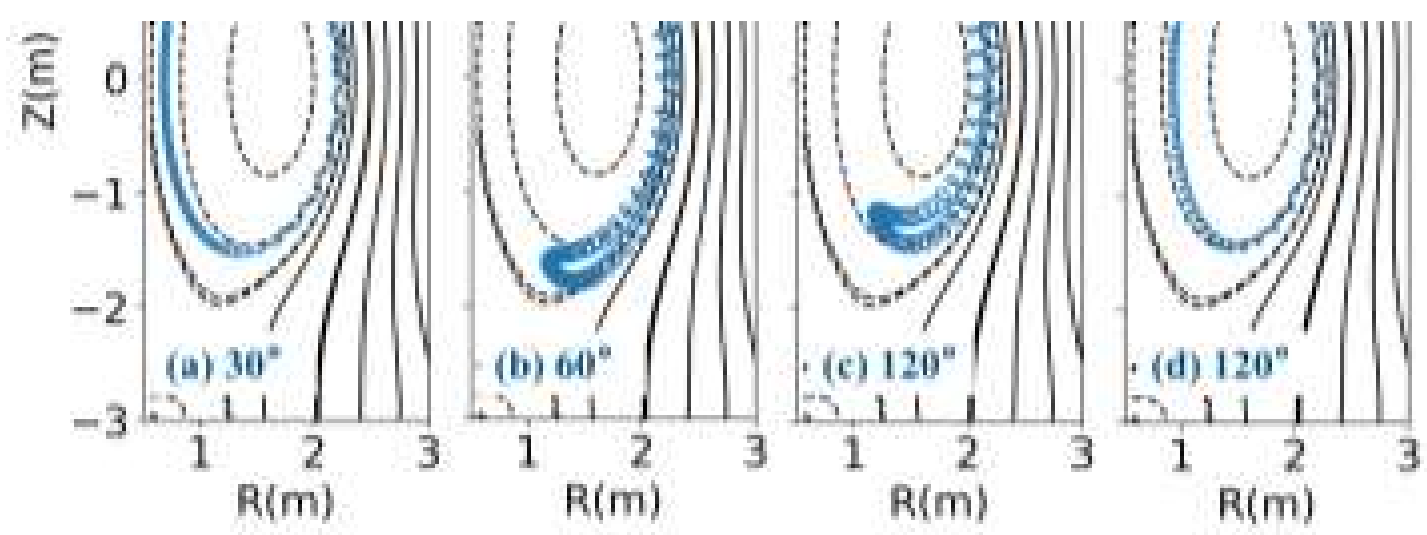

**Fig. 6.** Orbit of a 3.5 MeV proton starting from the mid-plane outboard near the LCFS with an initial velocity at an angle of (a) 30°, (b) 60°, (c) 120°, and (d) 150° relative to the local magnetic field. Panels (a) and (b) represent counter-current protons, and panels (c) and (d) co-current protons.

Figure 6 shows representative orbits for 3.5 MeV protons launched near the outboard mid-plane at the LCFS edge, for initial velocity angles of (a) 30°, (b) 60°, (c) 120°, and (d) 150° relative to the local magnetic field. The counter-current orbits drift outward, sampling higher-density thermal boron ions than at their launch point on the mid-plane, and vice versa for the counter-current proton orbits. The effective yield, therefore, depends on the phase-space magnitude of co- and counter-current passing protons.

As indicated by Figure 13 of [28], co-current bulk rotation increases the number of co-current passing orbits that drift inward; counterrotation increases outward-drifting passing orbits. This non-local orbit effect is absent from an unmagnetized 0D model in which reactivity depends only on local distributions and relative drift velocities [50]. Consequently, the local 0D treatment used to obtain Figure 5 does not adequately describe the ST $p-^{11}B$ plasma of Section 3.

Orbit loss also modifies the reacting population near the LCFS. Outward-drifting counter-current suprathermal protons launched near the edge can intersect material boundaries and be lost before completing a confined orbit, depleting the reacting population and reducing fusion contributions relative to the local-model estimate.

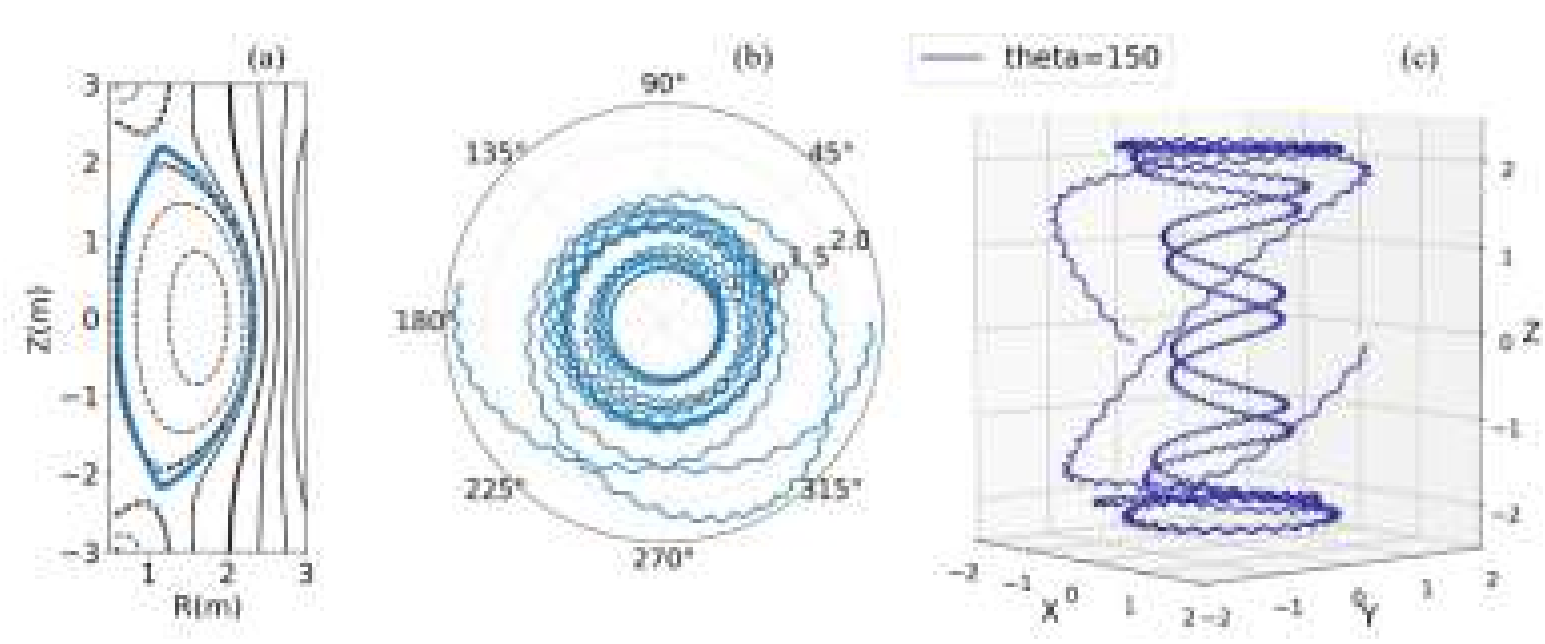

**Fig. 7.** (a) Poloidal, (b) top-down, and (c) 3D views of an example co-current drift orbit of a suprathermal proton with energy 2.13 MeV and pitch angle 150°, starting from the outer mid-plane position $R = 2.31$m just within the LCFS. The orbit begins and ends on the mid-plane and makes three additional transits through the mid-plane.

Figure 7 provides an example of a contained co-current passing orbit of a suprathermal proton with energy 2.13 MeV and pitch angle 150°, launched from the outboard mid-plane at $R = 2.31$m just inside the LCFS. The orbital trajectory is longer in the inboard region, where thermal boron density is higher than at the launch position, thereby increasing the orbit-averaged reaction probability. Figure 8 maps contained and lost orbit regions in phase space for three launch radii near the outboard LCFS, showing that orbit loss can remove a substantial fraction of suprathermal ions within the outer ~10 cm in major radius. As a result, local fusion power estimates, such as those shown in Figure 5, require correction.

### 4.3. Non-local computation of fusion reaction rate in a burning ST p-$^{11}$B plasma

To compute suprathermal-proton fusion power consistently with the equilibrium of Section 3, the reaction probability should be integrated along suprathermal proton drift orbits, averaged in velocity space, and then integrated in real space over the orbit-launch locations along the mid-plane. The computation proceeds as:

1. **Define launch positions.** Parameterize the outboard mid-plane by major radius $R$ from the magnetic axis to the outer edge of the orbit-containment region with $Z = 0$ and $\phi = 0$, and sample at equal spacing.

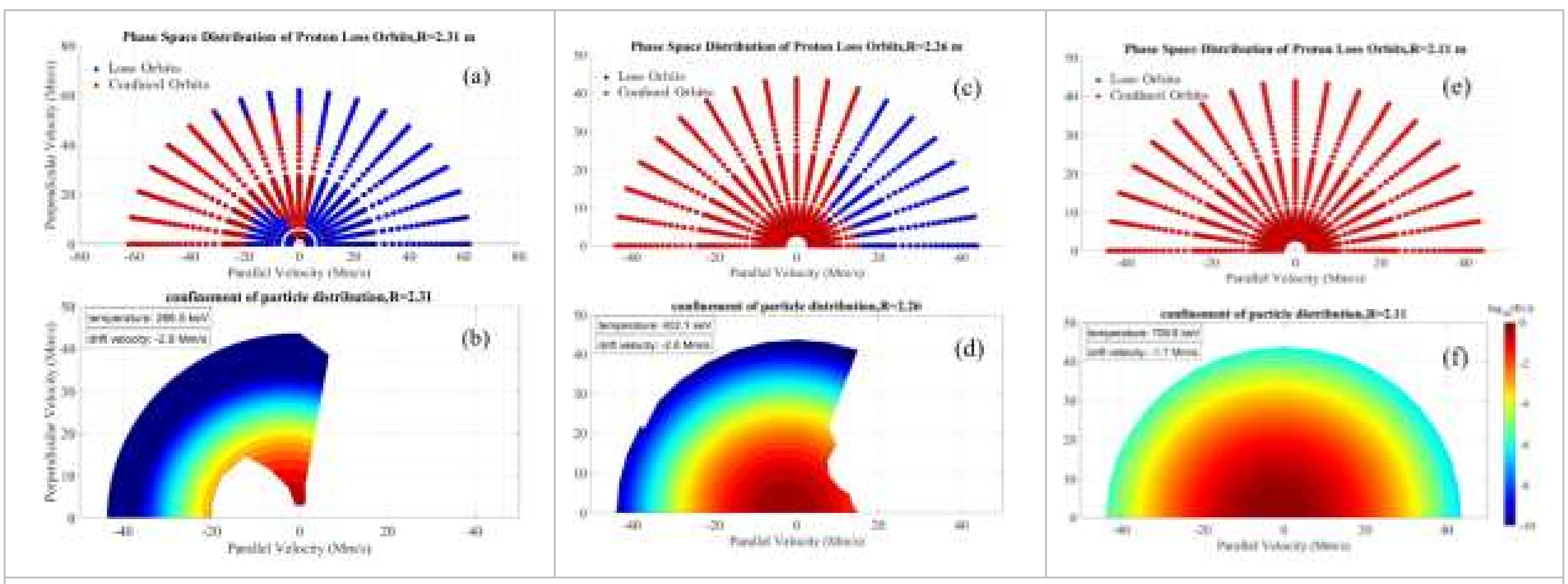

Fig. 8. Phase-space region of contained and lost orbits of suprathermal protons starting from the outboard mid-plane: (a) - (b) up to $6 \times 10^7$ $\mathrm{m\,s^{-1}}$($\sim 23$ MeV) from $R = 2.31$m; (c) - (d) and (e) - (f) up to $4 \times 10^7$ $\mathrm{m\,s^{-1}}$($\sim 10$ MeV) from $R = 2.26$m and $R = 2.21$m, respectively. Approximate phase-space regions of contained orbits are shown over a backdrop of a Maxwellian velocity distribution in (b), (d), and (f).

2. **Sample initial velocities.** For convenience, take thermal $^{11}$B ions and suprathermal protons to satisfy drifting Maxwellian distributions

$$f(v_R, v_\phi, v_Z) = n\left(\frac{m}{2\pi T}\right)^{3/2} \exp\left[\ -\frac{m}{2T}\left((v_R - u_R)^2 + (v_\phi - u_\phi)^2 + (v_Z - u_Z)^2\right)\right],$$

where $v_i$ and $u_i$ $(i = R, \phi, Z)$ are particle and fluid velocities, respectively, and sample from these distributions in velocity space.

3. **Track suprathermal proton trajectories.** Track drift trajectories in the equilibrium fields; compute $\sigma(|\mathbf{v}_1 - \mathbf{v}_2|)\,|\mathbf{v}_1 - \mathbf{v}_2|$ against the background thermal $^{11}$B distribution along the orbit.

4. **Orbit-average reaction probability.** Average the reaction probability over a complete orbit segment that nearly returns to the launch mid-plane location.

5. **Accumulate over velocity space and laboratory space.** Sum orbit-averaged contributions over the launch velocities and positions, weighted by density profile and pitch-angle distribution of the suprathermal protons on the mid-plane.

Drift-orbit deviation from flux surfaces is relatively small for the remaining reacting ion combinations, so their fusion contributions can be computed using the conventional local model. In contrast, the suprathermal-proton contribution must be treated non-locally to capture the inward-drift sampling of higher boron density, preferential co-current passing orbit effects due to toroidal plasma rotation, and orbit losses near the LCFS (Figures 8).

It is helpful to note that the non-local effects of suprathermal trapped orbits include both the co-current and counter-current components, similar to the relativistic trapped electron orbits described by Maekawa [41]. A proton orbit launched from the outboard mid-plane in the co-current direction becomes a counter-current orbit after reflecting from the "banana tip", and vice versa. Initially co-current orbits invariably extend into the plasma core, where they encounter significantly higher thermal boron densities, thereby increasing fusion reactivity, and vice versa. The fractions of the co- versus counter-launched trapped orbits increase with increasing plasma co-current rotational speed, as shown in Figure 8. A net effect of a strong co-rotation is therefore an increased fusion reactivity.

## 5. Implications and new physics features of burning ST p-$^{11}$B plasmas

The equilibrium and orbit results above are obtained from a multi-magnetofluid axisymmetric force-balance model benchmarked against EXL-50/EXL-50U equilibrium reconstruction [11] and extended here to an ST p-$^{11}$B plasma with strong rotation and minority suprathermal components (Section 3). Section 4 further shows that suprathermal proton fusion contributions can be strongly non-local in a compact ST because of orbit topology, the preferential inward drift of co-current passing ions, and orbit loss near the LCFS, which can substantially modify the effective reaction probability relative to a local Maxwellian estimate. While no p-$^{11}$B-burning ST database exists yet, it is useful to outline how the same equilibrium and orbit properties may affect confinement physics beyond the reactivity correction discussed in Section 4.

Several topics deserve discussion. First, the predicted equilibria contain multiple-temperatures: thermal ions near the lower p-$^{11}$B cross-section peak coexist with minority suprathermal ions nearer the higher-energy peak, and with a relativistic electron component used for current sustainment (Section 3). Second, the resulting magnetostatic and electrostatic fields are not independent background quantities; they are solved self-consistently so that each component fluid satisfies force balance simultaneously. Third, in low-aspect-ratio, high-beta ST equilibria, orbit widths and orbit topology can be integral parts of the macroscopic state, rather than small corrections, and can therefore affect transport, stability, and edge conditions in ways not anticipated in a thermonuclear plasma such as described in the ITER physics basis [42].

### 5.1. Reversed $\boldsymbol{E}\times\boldsymbol{B}$ shear and toroidal precession, and an absolute well in $|\boldsymbol{B}|$

Transport barriers associated with $E\times B$ flow shear have been observed across a wide range of tokamak regimes, including internal transport barriers (ITBs) in strongly rotating discharges [43,56,57]. In the projected ST p-$^{11}$B equilibrium of Section 3, strong toroidal rotation is accompanied by significant shear localized within an outboard magnetic-well region (Figures 3 and 4). This alone suggests that a barrier-like reduction in turbulent transport is likely, but the present configuration introduces additional features.

A notable feature in these equilibria is the simultaneous appearance of (i) reversed $E \times B$ shear throughout the core and (ii) reversed drift-orbit precession arising from the combination of high beta and low aspect ratio. In conventional tokamak discussions, these two ingredients are usually treated separately; $E \times B$ shear introduces turbulent decorrelation and stabilization, while precession introduces energetic-particle resonances and microinstabilities. When both reverse in tandem, the resulting shear landscape experienced by ion-scale and electron-scale turbulence can differ materially from expectations in a thermonuclear D-T plasma [42].

Electron-scale behavior is especially relevant here because p - $^{11}$B burn favors $T_i > T_e$ to reduce radiation losses while keeping ions close to the p - $^{11}$B cross-section peaks. That aim, together with strong rotation, places a premium on suppressing electron temperature gradient (ETG)-driven energy transport. The combined reversal described above points to the possibility of electron thermal internal transport barrier formation driven by shear and precession effects, as analyzed under a similar theoretical condition [57]. It also connects to the broader point that energetic populations can modify turbulence: recent JET D - T work has emphasized that highly energetic ions can influence turbulence suppression in ways that require global modelling [58]. The ST p - $^{11}$B state considered here is, by construction, rich in suprathermal populations, implying that turbulence models should be tested against equilibria containing those components rather than against purely thermal Maxwellian assumptions.

In addition, the equilibrium exhibits an absolute magnetic well in $|B|$in the outboard region (Figure 3e), associated with axisymmetric omnigeneity [48,49] and consistent with early ST feature calculations [47]. Such a $|B|$ well can, in principle, reduce the drive for certain trapped-particle-related transport channels and increase stability margins for electron-scale modes. Combined with the strong rotation and shear, this raises the prospect that confinement improvement beyond existing compact high-field ST experiments could be realized if impurity control and power balance are maintained simultaneously. The recent high-field compact ST results, including ST40 operation reaching central ion temperatures of $10^8$K [8] and Globus-M2 scaling showing strong sensitivity to toroidal field [7], provide experimental motivation for this line of reasoning, although those devices operate in deuterium/hydrogen, not in a p - $^{11}$B.

Finally, it is important to note that in this paper, the absolute well and reversed-shear features are equilibrium outputs, not imposed inputs. They follow from the multi-magnetofluid force balance with strong flow and finite suprathermal pressure contributions. That is a key difference relative to postulated transport barriers that are introduced phenomenologically. Here, the equilibrium itself provides a structured environment in which transport barriers could form and a well-defined starting point for gyrokinetic and neoclassical analysis.

### 5.2. Reformulation of neoclassical transport for magnetic-well, strong-flow ST plasmas

Standard neoclassical transport theory in toroidal confinement systems provides baseline scaling expectations for ion and electron transport as a function of collisionality, orbit classification, and magnetic geometry [44]. For conventional aspect ratios and moderate flow, the separation between equilibrium (MHD) and neoclassical correction is often workable: one computes equilibrium, then evaluates neoclassical transport on the resulting flux surfaces. The configuration described here challenges that separation in two ways: the equilibrium intentionally forms a large outboard $|B|$ well and omnigenous region (Figure 3e), and the ion population includes minority suprathermal components with orbit widths that are not negligible fractions of equilibrium gradient scale lengths (Section 4).

Shaing and collaborators developed transport theory extensions to include potato orbits in an axisymmetric torus with finite toroidal flow speed [51]. Later, they proposed that improved neoclassical confinement and turbulence suppression can occur in magnetic-well configurations relevant to reactor regimes [52]. These works provide a conceptual pathway to interpret the present equilibrium: an outboard well and strong rotation can squeeze orbits, reduce radial excursions, and thereby reduce neoclassical losses. This is consistent with the earlier claim in the manuscript that ion energy confinement may scale more favorably with $T_i$ in such a well region than with standard expectations.

The present ST p-$^{11}$B scenario adds a further twist: the non-local orbit behavior of suprathermal protons (Section 4) implies that even the meaning of the local transport coefficient can be ambiguous near the LCFS and in the outboard region of the plasma. If a significant fraction of suprathermal ions is orbit-contained but samples a broader radial range, then the transport description may require orbit-averaged fluxes rather than flux-surface-local averages. Conversely, if orbit loss is large in the outer ~10 cm (Figure 8), then the suprathermal distribution function can become strongly non-Maxwellian and anisotropic, again undermining assumptions that enter many neoclassical derivations.

Another ingredient that can force reformulation is the electrostatic potential. The equilibrium produces a positive $V_E$ relative to the wall (Figure 3d). Even if one focuses on neoclassical ion transport, $V_E$ influences trapped-particle fractions, orbit turning points, and the $E \times B$ drift, all of which enter neoclassical flux calculations. In a multi-fluid equilibrium, $V_E$ is a consequence of the multi-magnetofluid equilibrium solution, which provides a new basis for neoclassical transport.

For these reasons, a consistent neoclassical description for this regime should incorporate: (i) the computed $|B|$ structure and omnigenous region, (ii) strong toroidal flow and shear, (iii) the self-consistent electrostatic potential profile, and (iv) multi-temperature ion populations including suprathermal minorities. This paper does not provide a new neoclassical theory; rather, it identifies the physical regime in which standard neoclassical predictions may be systematically misleading and motivates new modelling to be validated against new data from EXL-50U and the upcoming EHL-2 [45,46].

### 5.3. Suppression of tearing modes and macroscopic stability

Macroscopic stability and operational limits remain central constraints for any high-beta ST concept. The ITER MHD stability basis summarizes broad classes of limits and the role of equilibrium profiles in triggering or avoiding disruptive behavior [39], but ST operation introduces additional factors such as strong rotation and close-fitting conducting structures. In NSTX, for example, rotation played a key role in stabilizing resistive wall modes in high-beta plasmas [5]. The ST p-$^{11}$B equilibrium of section 3 is designed to operate at elevated $\beta$ with substantial rotation; therefore, the NSTX experience suggests that rotation may widen the stable operating space in such a case.

However, the equilibrium structure here differs from standard MHD equilibria in ways that could affect tearing and related resistive instabilities. The current profile includes a significant suprathermal electron contribution that can extend beyond the LCFS (Figure 3c), which is atypical of tokamak equilibria, where current is largely limited to within the LCFS. Even though the equilibrium is axisymmetric, the extension of current beyond LCFS implies that the edge and scrape-off-like regions can carry active current and potentially modify rational-surface structure and tearing drive in the outboard peripheral region. In addition, in a multi-magnetofluid equilibrium, the relationship among pressure gradients, flow shear, and current-density gradients may differ from that in a single-fluid MHD model, especially when suprathermal pressures and relativistic-electron dynamics are present.

Another stability issue arises from the presence of suprathermal ions and electrons. Such populations can interact with MHD modes via resonances, provide drive or damping depending on distribution anisotropy, and alter the effective moment of inertia and rotation profile. While detailed kinetic-MHD analysis is beyond the scope of this paper, the appearance of strong differential rotation, e.g., velocity differences exceeding $2\times10^{6}\ \mathrm{m\,s^{-1}}$between suprathermal protons and thermal boron ions (Section 3), suggests that shear-driven stabilization of some modes may coexist with new free-energy sources for others.

Therefore, suppression of tearing modes in this regime should be treated as a research question rather than a guaranteed benefit. The equilibrium solver provides quantitative profiles for current, pressure, flow, and $V_E$. These profiles can be used as inputs for stability analysis consistent with tokamak formulations [39] and with prior ST equilibrium models incorporating suprathermal ions [40]. In this sense, the present paper's main contribution is to supply a physics-motivated new equilibrium state in which such stability questions can be posed for a p-$^{11}$B ST fusion plasma.

### 5.4. Positive plasma potential and implications for alpha ash and impurity exhaust

In the high-beta p-$^{11}$B equilibrium example, the computed electrostatic potential reaches a peak value of ~10.8 kV relative to the wall (Figure 3d). This potential is not an arbitrary assumption; it emerges from the requirement that each species satisfies force balance under common fields in the multi-magnetofluid model. A positive plasma potential influences particle confinement in several ways. Low-energy ions that diffuse or drift beyond the LCFS are electrostatically discouraged from reentering the peripheral plasma. At the same time, suprathermal populations can extend beyond the LCFS due to their orbital widths, and the potential can modify where they turn, how they sample densities, and how they interact with nearby material surfaces.

For p-$^{11}$B plasmas, ash management is a central issue because the fusion products are alpha particles. In advanced tokamak scenarios, impurity and ash accumulation have been identified as a key uncertainty that can terminate high-performance phases [56]. In the ST p-$^{11}$B case, the positive $V_E$ and the altered orbit structure may, in principle, permit additional mechanisms for ash and impurity control. For example, if the potential and rotation produce favorable drift patterns for helium ions, the residence time of ash in the core could be reduced. Alternatively, if ash becomes trapped in certain orbit classes or if edge barriers become too effective, accumulation could worsen. The sign and magnitude of these effects depend on collisionality, turbulence, and the detailed edge and divertor/limiter configuration.

Another practical implication is that a positive potential may alter the balance between sputtering and impurity source terms on plasma-facing components, particularly when energetic electrons or ions impinge on the surfaces. In EXL-50U reconstructions, a large fraction of current is inferred to be carried by energetic electrons whose orbits can extend beyond the LCFS (Section 2). In the projected burning p-$^{11}$B plasma, suprathermal electrons likewise extend beyond LCFS and carry about 21% of total current (Section 3). This raises a coupled engineering-physics question: whether the same potential structure that supports force balance can be maintained while also keeping wall power loading and impurity production within acceptable bounds.

Thus, the positive plasma potential is likely both an opportunity and a constraint. It could contribute to improved core confinement and edge control, but it also emphasizes the need for integrated modelling: equilibrium, orbit loss, wall interaction, and impurity transport must be treated together for credible predictions.

### 5.5. Raised pedestal height via recycling reduction and modified edge conditions

The equilibrium results show that suprathermal ions and electrons can extend well beyond the LCFS and be limited by the nearest plasma-facing component (Figures 1 and 3). This structure suggests that edge conditions may not resemble conventional H-mode pedestal physics in tokamaks, where a sharp separatrix and SOL dynamics dominate recycling and pedestal fueling. Here, the energetic populations can reach the wall or limiter directly through orbit excursions, and the plasma may form a broader transition region between the thermal-plasma LCFS and the material boundary for the suprathermal components.

One likely consequence is a change in recycling. If the edge is partially depleted of cold ions and neutrals because the positive $V_E$ repels low-energy ions and because the energetic component interacts differently with surfaces, the effective neutral source penetrating the core would decrease. Reduced neutral fueling at the LCFS can, in some regimes, permit higher temperatures and pressure gradients at the LCFS, and raise an effective pedestal height. Such a mechanism would be beneficial for overall confinement, particularly if it couples to the strong shear region in the outboard magnetic well region (Section 5.1). However, recycling is expected to become more complex due to orbit geometry and the loss of suprathermal particles.

The orbit-loss maps in Figure 8 highlight that a substantial fraction of suprathermal protons passing near the outer ~10 cm can be lost. These losses can create localized wall interactions, prompt charge-exchange neutrals, and produce particle sources that are spatially and energetically distinct from those in standard SOL models. Therefore, an optimistic recycling-reduction picture must be balanced against the possibility that the suprathermal-orbit loss becomes a dominant source of edge particle and power flux densities.

In addition, the edge and pedestal behavior can be sensitive to non-local orbit sampling. As Section 4 emphasizes, co-current passing suprathermal protons can drift inward and sample higher thermal-boron density. This raises the possibility that a suprathermal component could contribute to self-heating in regions offset from its density peak, altering local gradients and potentially affecting pedestal formation indirectly. In this sense, pedestal height in such a plasma is not controlled solely by edge-local physics, but by the global orbit structure of the suprathermal ions.

The above motivates two modelling needs. First, coupling equilibrium/orbit calculations to edge neutral and recycling models is needed. Second, experimental characterization of edge conditions in EXL-50U and the upcoming EHL-2 [40,41] should include diagnostics sensitive to both thermal edge parameters and energetic-particle losses, since both can control the effective pedestal and, by extension, global confinement.

### 5.6. Reduction of non-axisymmetric fields to allow high rotation and preserve omnigeneity

The equilibria and the performance benefits proposed in this paper rely heavily on strong toroidal rotation and on the formation of an outboard omnigenous region linked to the $|B|$ well (Figure 3e). In an axisymmetric description, omnigeneity can reduce certain classes of neoclassical transport [48,49]. However, non-axisymmetric perturbations—error fields, coil ripple, structural asymmetries—can break omnigeneity, drive additional transport of the suprathermal plasma components, and degrade rotation through neoclassical toroidal viscosity. Therefore, the engineering tolerance for three-dimensional field components may be more stringent in this scenario than in thermal-plasma-only ST discharges.

High rotation also interacts with stability. In NSTX, rotation assisted in maintaining stability at high beta [5]. For the present p-$^{11}$B plasma, a similar reliance on rotation implies that uncontrolled non-axisymmetric fields that slow rotation or create resonant braking can shrink the accessible operating space. This risk is heightened if the projected regime approaches very high Mach numbers or super-Alfvénic conditions in the outboard region, as strong rotation has been observed in some ST regimes cited in the Introduction.

In addition, because suprathermal electrons can carry a substantial fraction of current and can extend beyond LCFS (Sections 2 and 3), non-axisymmetric fields could also modify their orbit topology, alter where they form current, and change the inferred current profile. This would affect MHD stability and the equilibrium itself. Therefore, maintaining high-quality axisymmetry is not only a transport issue but also an issue of equilibrium maintenance of the multi-magnetofluid plasma state.

Pragmatically, this motivates design and operational priorities for future experiments aiming at the p-$^{11}$B regime: careful error-field correction, minimizing localized ferromagnetic perturbations, and controlling externally applied non-axisymmetric fields that could compromise rotation and orbit containment of the suprathermal ions. If these requirements are satisfied, it may permit sustained high rotation while preserving the equilibrium features—$|B|$ well, strong $E \times B$ shear, and non-local orbit containment—on which the proposed confinement and reactivity improvements depend.

---

## 6. Summary and conclusions

A rotating ST p-$^{11}$B plasma concept has been formulated using a multi-component magnetofluid force-balance model, extending earlier work validated by the EXL-50/EXL-50U equilibrium reconstruction, which includes relativistic electrons [11].

The present model targets plasmas that combine (i) thermalized bulk ions and electrons with ion temperatures exceeding 100 keV and densities above $10^{20}\ \mathrm{m}^{-3}$, and (ii) minority suprathermal ions and electrons, with ion energies in the ~0.5 MeV range and electron energies in the MeV range. This combination is motivated by the p-$^{11}$B fusion cross section, which has a local peak near 160–165 keV and a broader, higher peak near 650–675 keV [1,2].

The paper identifies and emphasizes several coupled equilibrium-orbit features that may be advantageous for p-$^{11}$B fusion plasma operation in a compact ST geometry.

1. **Component-wise force balance under common fields.** Each species satisfies a force balance including centripetal, electrostatic, and Lorentz terms under the same $\mathbf{E}$ and $\mathbf{B}$ fields, leading to species-dependent rotation and strong differential flow speeds. In the near-burn example equilibrium, the velocity difference between suprathermal protons and thermal boron ions exceeds $2 \times 10^6\ \mathrm{m\,s^{-1}}$ over a substantial region, with peak relative speeds near 2700 km s $^{-1}$ at the LCFS.
2. **High-beta equilibria with an outboard $|B|$ well and axisymmetric omnigeneity.** The computed equilibria exhibit an outboard magnetic well-and-hill structure in $|B|$ consistent with early ST feature expectations [47] and with omnigeneity concepts [48,49]. Such a region is expected to modify the orbit topology (via orbit squeezing) and may reduce neoclassical losses in a manner consistent with early developments in the magnetic-well transport theory [51,52].
3. **Suprathermal components extending beyond the LCFS.** In both EXL-50U reconstructions and in the projected burning p-$^{11}$B equilibria, energetic electrons can extend beyond the LCFS. In the burning-plasma equilibrium, the suprathermal electron component carries $\approx 21\%$ of the total plasma current and extends beyond the LCFS. A positive electrostatic potential is obtained self-consistently, with a peak value of ~10 kV relative to the wall, thereby affecting edge confinement of low-temperature ash and impurity ions.
4. **Non-local fusion reactivity for suprathermal protons.** A local (0D) drifting-Maxwellian reactivity model predicts a volume-averaged fusion power density $P_{pB} = 0.0306\ \mathrm{MW\,m^{-3}}$ for the example equilibrium, with the suprathermal-proton/thermal-boron ion reaction contributing only 8.8% under local fusion assumptions. Orbit calculations, however, show that this local estimate can be substantially inaccurate because MeV-level suprathermal protons have drift orbits that sample wide radial regions; co-current passing orbits drift inward into higher boron density, while counter-current orbits drift outward and can be lost if they strike the plasma-facing surface. A consistent evaluation, therefore, requires non-local orbit integration to determine the reaction probability, followed by velocity-space and real-space averaging, as outlined in Section 4.

5. **Motivation for new confinement and stability studies.** The combined results motivate confinement and stability research for p-$^{11}$B ST plasmas that differ from the thermonuclear ITER-like regime summarized in the ITER physics basis [42]. In particular, the computed equilibria imply reversed shear structures and large $|B|$ wells that may alter microturbulence and neoclassical transport. At the same time, the characteristic plasma current and suprathermal particle distributions raise questions about stability and their impact on plasma-facing components, which require dedicated analysis and experimental validation.

Finally, the work highlights the need for new experiments to build a database that extends beyond the parameters of EXL-50U toward those proposed for burning ST p-$^{11}$B plasmas. The upcoming EHL-2 ST under preparation targets the parameter space of plasma current $I_p \sim 3$MA, major radius $R \sim 1$m, and toroidal field $B_T \sim 3$T [45,46], to increase ion temperatures toward a few tens of keV along the path of ENN's p-$^{11}$B fusion roadmap [12]. This will enable systematic tests of the projected multi-magnetofluid, double-temperature, strongly rotating, high-toroidal-beta plasma equilibrium; suprathermal ion and electron orbit behavior with the non-local suprathermal-to-thermal fusion reaction rates; potential confinement and stability improvements, with the prospect of reduced particle recycling at the LCFS at the expense of increased wall impurity sources; and current-drive physics, including the dominating ion current far exceeding the total plasma current, likely reversing the direction of the thermal electron current.

This work also highlights the challenge due to the much lower p-$^{11}$B fusion power density compared to D-T fusion, assuming similar plasma densities. This implies that the overall confinement time of such a plasma must be much longer than that dominated by the well-known microturbulences of today's toroidal confinement devices, further underscoring the importance of the upcoming EHL-2 experimental research plans. Of equal importance is the complex field of p-$^{11}$B fusion reaction physics, which remains open to fundamental issues of understanding and possible manipulation [59].

---

**Acknowledgements**

Discussions on a range of topics with Drs. Feng Kaiming, Liang Yunfeng, Xie Huasheng, Dong Jiaqi, and Yang Yuanming are gratefully acknowledged.

---